\documentclass[aps,11pt]{revtex4}
\usepackage[dvips]{graphicx}

\begin{document}

\title{Spin-orbit origin of large reduction of the effective moment in Na$_{2}$V$_{3}$O$_{7}$$^+$}
\author{R.J. Radwanski}
\homepage{http://www.css-physics.edu.pl} \email{sfradwan@cyf-kr.edu.pl}
\affiliation{Center for Solid State Physics, S$^{nt}$Filip 5, 31-150 Krakow, Poland,\\
Institute of Physics, Pedagogical University, 30-084 Krakow, Poland}
\author{Z. Ropka}
\affiliation{Center for Solid State Physics, S$^{nt}$Filip 5, 31-150 Krakow, Poland}

\begin{abstract}
We have shown that the observed large reduction of the effective moment and the drastic violation of the Curie-Weiss
law in Na$_{2}$V$_{3}$O$_{7}$ (Phys. Rev. Lett. 90 (2003) 167202) is caused by conventional crystal-field interactions
and the intra-atomic spin-orbit coupling of the V$^{4+}$ ion. The fine discrete electronic structure of the 3$d^{1}$
configuration with the weakly-magnetic Kramers-doublet ground state, caused by the large
orbital moment, is the reason for anomalous properties of Na$_{2}$V$_{3}$O$%
_{7}$. Moreover, according to the Quantum Atomistic Solid-State Theory (QUASST) Na$_{2}$V$_{3}$O$_{7}$ is expected to
exhibit pronounced heavy-fermion phenomena at low temperatures.


\end{abstract}

\maketitle

Gavilano {\it et al.} \cite{1} have discovered that the effective moment of the V$^{4+}$ ion in Na$_{2}$V$_{3}$O$_{7}$
is reduced by the one order of magnitude upon reducing the temperature from 100 to 10 K. The moment reduction has been
inferred from the experimentally measured temperature dependence of the magnetic susceptibility. In the figure 1 of
Ref. \cite{1} it is seen that after taking into account the diamagnetic contribution $\chi _{o}$ the inverse
susceptibility shows in the temperature range 100-300 K a straight line behavior with the effective moment p$_{eff}$ of
1.9 $\mu _{B}$ per V ion. Another straight line between 20 and 1.9 K implies $p_{eff}$ of one order of magnitude
smaller. Gavilano {\it et al.} provide an explanation that ''The reduction of the effective magnetic moment is most
likely due to a gradual process of moment compensation via the formation of singlet spin configurations with most but
not all of the ions taking part in this process. This may be the result of antiferromagnetic interactions and
geometrical frustration.'' They further conjectured ''the compensation of eight out of the nine V spins ...'' in order
to reproduce the observed
reduction of the effective moment by one order of magnitude. Na$_{2}$V$_{3}$O%
$_{7}$ shows no sign of the magnetic order down to 1.9 K.

The aim of this Letter is to propose a more physical explanation for this reduction of the effective moment in
Na$_{2}$V$_{3}$O$_{7}$. We can reproduce very well the observed temperature dependence of the paramagnetic
susceptibility, remarkably violating the Curie-Weiss law, by considering the electronic structure associated with the
V$^{4+}$ ion (3$d^{1}$ configuration) under the action of the crystal field (CEF)\ taking into account the spin-orbit
(s-o) coupling. Despite of the Kramers doublet ground state, a ground-state with quite small magnetic moment can be
obtained as an effect of the spin-orbit coupling. It turns out that even weak s-o coupling unquenches a quite large
orbital moment. In our present explanation we have been oriented by our earlier calculations for the V$^{4+}$ ion
presented in Refs \cite{2,3}.

The one 3$d$ electron in the V$^{4+}$ ion is described by quantum numbers $L$%
=2 and $S$=1/2. The ground term $^{2}D$ is 10-fold degenerated. Its degeneracy is removed by the intra-atomic
spin-orbit interactions and in a solid by crystal-field interactions. This situation can be exactly traced by the
consideration of a single-ion-like Hamiltonian

\begin{center}
$H_{d}=H_{CF}^{octa}+H_{s-o}+H_{CF}^{tr}+H_{Z}=B_{4}(O_{4}^{0}+5O_{4}^{4})+%
\lambda L\cdot S+B_{2}^{0}O_{2}^{0}+\mu _{B}(L+g_{e}S)\cdot {\bf B}\,\;\;\;$
\end{center}

in the 10-fold degenerated spin-orbital space. The last term allows to calculate the influence of the external magnetic
field {\bf B} and enables calculations of the paramagnetic susceptibility. Such type of the single-ion Hamiltonian has
been widely used in analysis of electron-paramagnetic resonance (EPR)\ spectra of 3$d$-ion doped systems \cite{4,5}.
Here we use this Hamiltonian for systems, where the 3$d$ ion is the full part of the crystal. Although we know that the
local symmetry in Na$_{2}$V$_{3}$O$_{7}$ is quite complex we approximate, for simplicity, the CEF\ interactions at the
V site by considering dominant octahedral interactions with a trigonal distortion. For the octahedral crystal field we
take $B_{4}$= +200 K. The sign ''+'' comes up from {\it ab initio} calculations for the ligand octahedron \cite{6}. The
spin-orbit coupling parameter $\lambda _{s-o}$ is taken as +360 K, as in the free V$^{4+}$ ion \cite{4}.

The resulting electronic structure of the 3$d^{1}$ ion contains 5 Kramers doublets separated in case of the dominant
octahedral CEF interactions into 3 lower doublets, the $T_{2g}$ cubic subterm, and 2 doublets, the $E_{g}$ subterm,
about 2 eV above (Fig. 1). The $T_{2g}$ subterm in the presence of the spin-orbit coupling is further split into a
lower quartet and an excited
doublet (Fig. 1(2)). Positive values of the trigonal distortion parameter $%
B_{2}^{0}$ yields the ground state that has a small magnetic moment (Fig.
1(3)). For $B_{2}^{0}$ = +6 K the ground state moment amounts to $\pm 0.15$ $%
\mu _{B}$. It is composed from the spin moment of $\pm 0.42$ $\mu _{B}$ and the\ orbital moment of $\mp 0.27$ $\mu
_{B}$ (antiparallel). The sign $\pm $ corresponds to 2 Kramers conjugate states. The excited Kramers doublet lies at 38
K (3.3 meV) and is almost non-magnetic - its moment amounts to $\pm 0.03$ $\mu _{B}$ only (=$\pm 1.03$ $\mu _{B}+2\cdot
(\mp 0.50$ $\mu _{B})$) due to the cancellation of the spin moment by the orbital moment. So small and so different
moments for the subsequent energy levels is an effect of the spin-orbit coupling and distortions.
\begin{figure}[ht]
\includegraphics[width = 8.6 cm]{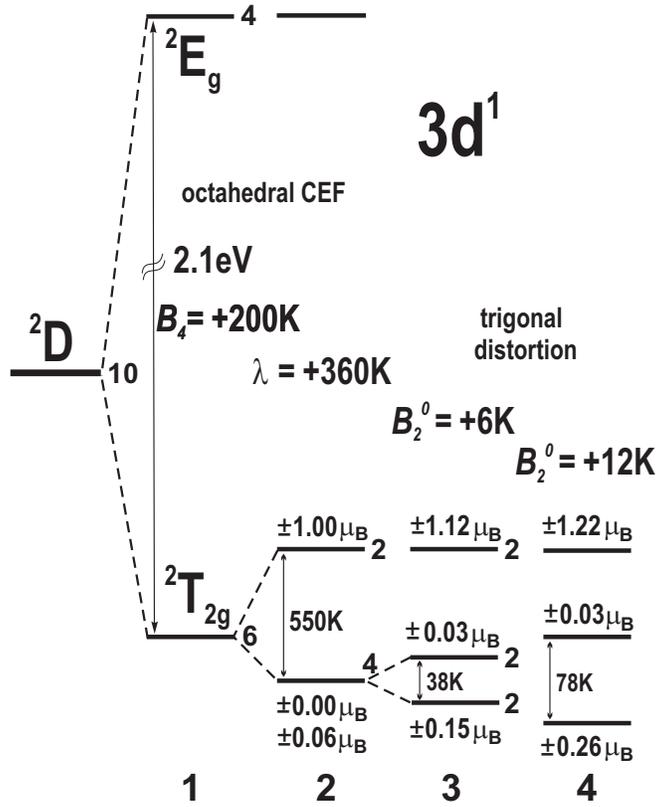}
\caption{Calculated localized states of the 3$d^{1}$ configuration in the V$%
^{4+}$ ion under the action of the crystal field and spin-orbit interactions originated from the 10-fold degenerated
$^{2}D$ term; (1) the splitting of
the $^{2}D$ term by the octahedral CEF\ surroundings with $B_{4}$=+200 K, $%
\lambda _{s-o}$ =0; (2) the splitting by the combined octahedral CEF and spin-orbit interactions; (3) and (4) the
effect of the trigonal distortions. The states are labelled by the degeneracy in the spin-orbital space and the value
of the magnetic moment.}
\end{figure}
\begin{figure}[ht]
\includegraphics[width = 11 cm]{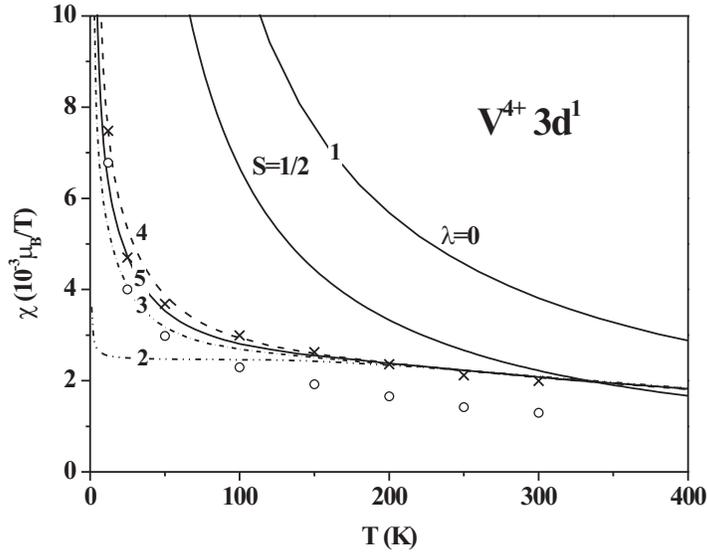}
\caption{The calculated temperature dependence of the atomic-scale paramagnetic susceptibility $\chi (T)$ for the
3$d^{1}$ configuration in the V$^{4+}$ ion for different physical situations: \ line (1) - $\chi (T)$ for the purely
octahedral crystal field with $B_{4}$=+200 K ($\lambda _{s-o}$= 0); line (2) - in combination with the spin-orbit
coupling $\lambda _{s-o}$= +360 K; lines (3) and (4) show the influence of the off-cubic trigonal
distortions $B_{2}^{0}$=+6 K (3) and $B_{2}^{0}$=+12 K (4); curve (5), with $%
B_{2}^{0}$=+9 K, reproduces very well measured experimental data (x) with taking into account the diamagnetic term
$\chi _{o}$ of -0.0007 $\mu _{B}$/T V-ion ($\simeq $ -0.0004 emu/mol V). Empty circles represent the rough experimental
data, after Refs \cite{1,7}.}
\end{figure}
\begin{figure}[ht]
\includegraphics[width = 11 cm]{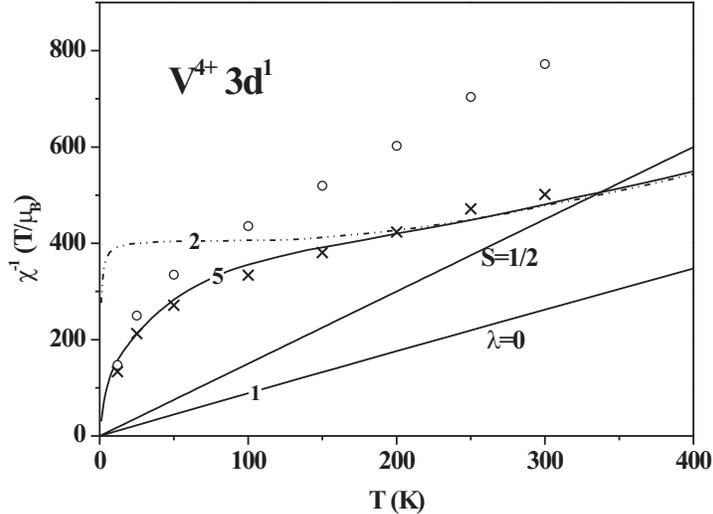}
\caption{The calculated temperature dependence of the atomic-scale paramagnetic susceptibility shown in the $\chi
^{-1}$ {\it vs} $T$ plot for the 3$d^{1}$ configuration in the V$^{4+}$ ion for different physical situations defined
in Fig. 1 and in the main text. The significant departure from the Curie law as well as from the S=1/2 behavior is an
effect of the spin-orbit coupling and distortions. The curve (5), with $B_{2}^{0}$=+9 K, reproduces very well measured
experimental data (x) with taking into account the diamagnetic term $\chi _{o}$ of -0.0007 $\mu _{B}$/T V-ion ($\simeq
$ -0.0004 emu/mol V)). Empty circles represent the rough experimental data, after Refs \cite{1,7}. The curves
corresponding to results (3) and (4) of Fig. 2 are not shown for clarity reasons.}
\end{figure}
\begin{figure}[ht]
\includegraphics[width = 10.2 cm]{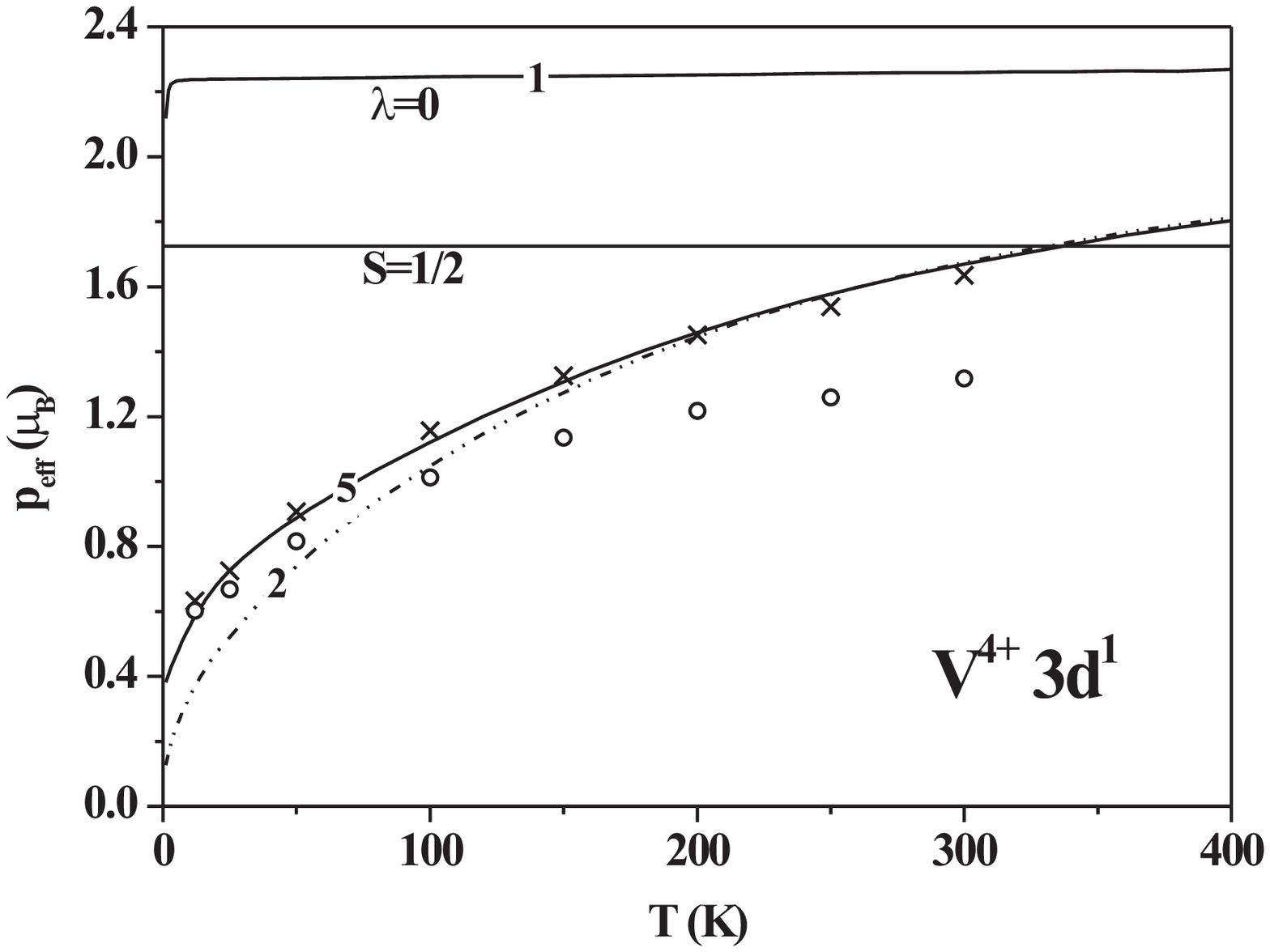}
\caption{Temperature dependence of the effective magnetic moment $p_{eff}$ calculated from the Curie law for the
3$d^{1}$ configuration in the V$^{4+}$ ion for different physical situations defined in Fig. 1 and in the main text.
The significant reduction from a value of 1.73 $\mu _{B}$, expected for the free spin S=1/2, is seen due to the
spin-orbit coupling and
distortions. The calculated curve (5), with $B_{4}$=+200 K, $\lambda _{s-o}$%
= 360 K and $B_{2}^{0}$=+9 K, reproduces very well measured experimental data (x) with taking into account the
diamagnetic term $\chi _{o}$ of -0.0007 $\mu _{B}$/T V-ion ($\simeq $ -0.0004 emu/mol V)). Empty circles represent
values obtained from the rough experimental data, after Refs \cite{1,7}.}
\end{figure}

Having the electronic structure, the state energies and eigenfunctions, we calculate the Helmholtz free energy
$F(T,B)$. Making use of the definition, known from statistical physics, we calculate the paramagnetic susceptibility,
as $\chi (T)$ $=$ $-\partial ^{2}F(T,B)/\partial B^{2}$, and its temperature dependence. In Figs 2 and 3 the calculated
results for the paramagnetic susceptibility are shown for different physical situations of the V$^{4+}$ ion: line (1) -
for the purely octahedral crystal field; line (2) shows $\chi (T)$ in the extra presence of the spin-orbit coupling;
lines (3), (4), (5) - in the extra presence of trigonal distortions $B_{2}^{0}>0$ of different strength (the case with
$B_{2}^{0}<0$ has been discussed in Ref. \cite{2} - in general it yields the temperature dependence of the
susceptibility with a broad maximum). The associated electronic structures are shown in Fig. 1. One can see that all
these curves are completely different from the Curie-law $S$=1/2 behavior, expected for the free one
spin. The V$^{4+}$ ion with one 3$d$ electron is usually treated as the $S$%
=1/2 system i.e. with the spin-only magnetism and with taking into account the spin degree of freedom only. The neglect
in the current literature of the orbital moment is consistent with the widely-spread conviction that the orbital
magnetism plays rather negligible role due to the quenching of the orbital moment for 3$d$ ions. Na$_{2}$V$_{3}$O$_{7}$
is an example of numerous compounds in which the $S$=1/2 behavior in the temperature
dependence of the paramagnetic susceptibility is drastically violated (CaV$%
_{4}$O$_{9}$, MgVO$_{3}$, (VO)$_{2}$P$_{2}$O$_{7}$, ...). This violation is seen in the substantial departure of $\chi
(T)$ from the Curie law at low temperatures. Our calculations prove that even weak spin-orbit coupling and small
distortions of the local surroundings of the V$^{4+}$ ion cause a rather drastic change of the slope of the $\chi
^{-1}(T)$ plot below 100 K (Fig. 3, lines (2) and (5)). Such change of the slope is usually attributed to the change of
the effective moment like it was made in Ref. \cite{1}. In Fig. 4 the temperature dependence of the effective moment
per the V ion, calculated as $p_{eff}=\sqrt{3k_{B}\chi T}$, k$_{B}$ denotes the Boltzmann constant, are presented for
different physical situations shown in Fig. 2. The strong temperature dependence is seen below 100 K, Fig. 4 lines (2)
and (5). Basing on these results we are convinced that the strong temperature dependence of the effective moment
inferred in Ref. \cite{1} results from the intra-atomic spin-orbit coupling and local distortions.

In Figs 2 and 3 we show the comparison of our calculated results with the experimental data, shown as x, taken from
Refs \cite{1,7} for the measured paramagnetic susceptibility and corrected for the diamagnetic term $\chi _{o} $ of
-0.0007 $\mu _{B}$/T V-ion ($\simeq $ -0.0004 emu/mol V). The corrected experimental data (x) coincide well with our
calculations shown as line 5 obtained for a set of parameters: $B_{4}$= +200 K, $\lambda _{s-o}$ = +360 K and
$B_{2}^{0}$ = +9 K. For this set of parameters the energy level scheme is the intermediate one to those shown as (3) or
(4) in Fig. 1. The energy level scheme contains 2 excited doublets at 58 \ and 580 K and the ground-state moment
amounts to 0.21 $\mu _{B}$. We treat this coincidence as
not fully relevant owing to the much more complex local symmetry of the V$%
^{4+}$ ion in Na$_{2}$V$_{3}$O$_{7}$, to a large uncertainty in the evaluation of the diamagnetic term and of the
paramagnetic susceptibility measured on a polycrystalline sample. We take, however, the reached agreement as strong
argument for the applied QUASST approach and strong
indication for the existence of the fine electronic structure in Na$_{2}$V$%
_{3}$O$_{7}$, originating from the V$^{4+}$ ion, determined by crystal-field and spin-orbit interactions. Extremely
important is a fact that our approach is able to reproduce the overall $\chi (T)$ dependence in the full measured
temperature range and that it reproduces the absolute value of the magnetic susceptibility.

In conclusion, we argue that the experimentally-observed temperature dependence of the paramagnetic susceptibility of
Na$_{2}$V$_{3}$O$_{7}$ with strong violation of the Curie-Weiss law can be explained in a single-ion
approach, in frame of the Quantum Atomistic Solid-State Theory QUASST \cite%
{8,9,10} for 3$d$-atom containing compounds, taking only into account conventional local atomic-scale effects like the
crystal-field interactions, the intra-atomic spin-orbit coupling and the orbital 3$d$ magnetism. The used parameters
$B_{4}$ (+200 K), $\lambda _{s-o}$ (+360 K) and $B_{2}^{0}$ (+9 K) have clear physical meaning. The used value of +200
K for $B_{4}$ is consistent with the value of +260 K found recently experimentally for the octahedral crystal field in
LaCoO$_{3}$ \cite{11} - the similarity is related to the fact that the strength of $B_{4}$ is predominantly determined
in both cases by the local oxygen octahedron. The obtained good reproduction of such nontrivial experimental results we
take as further indication for the high physical adequacy of the QUASST\ conjecture that the electronic and magnetic
properties of the 3{\it d-}ion containing compounds are predominantly determined by the fine electronic structure of
the 3{\it d} ion in the meV scale. Moreover, we would like to point out that according to QUASST the Kramers spin-like
degeneracy of the ground state, not removed down to 1.9 K, has to be removed somewhere - we expect it to occur at very
low temperatures. Thus QUASST predicts Na$_{2}$V$_{3}$O$_{7}$ to exhibit heavy-fermion-like properties in the specific
heat at ultra-low temperatures.\\

\noindent $^+$ This paper has been resubmitted 11-09-2003 to Phys.Rev.Lett., LG9202, after rejection 4-09-2003 of the
original submission of 10-07-2003 (cond-mat/0307272). The full documentation of the Phys.Rev.Lett. procedure as well as
two referee reports and our answers are collected as APPENDIX A-F. About the abnormal situation we have informed the
Editor-in-Chief and the President of the American Physical Society. The confining of the scientific discussion on just
published papers we treat as very serious violation of the scientific rules. We remind, we chose Phys.Rev.Lett. because
the paper of Gavilano et al., reporting properties of Na$_{2}$V$_{3}$O$_{7}$ with an exotic explanation, has appeared
in Phys.Rev.Lett. in April 2003 ({\bf 90}, 167202).

\newpage
\begin{center}
\textbf{APPENDIX A}\\
\textbf{\small{Submission letter to Physical Review Letters of 10-07-2003}}
\end{center}
To: \qquad prl@aps.org\\
Subject: \qquad submit Radwanski\\
Date sent: \qquad Thu, 10 Jul 2003 22:01:54 +0200\\
To Editor of Phys.Rev.Lett.

Reinhardt B. Schuhmann

Dear Editor,

Please find attached our new paper:

To the origin of large reduction of the effective moment in Na$_{2}$V$_{3}$O$%
_{7}$

by Z. Ropka and R.J. Radwanski, that we submit for publication in Phys. Rev. Lett.

The Phys.Rev.Lett. has been chosen as we provide novel physical explanation of large moment reduction in
Na$_{2}$V$_{3}$O$_{7}$ reported recently in Phys. Rev. Lett. {\bf 90} (2003) 167202. Thus our choice is fully
justified.

Paper is prepared in the RevTex.file. Paper accounts 6 pages with 8 references. It contains 4 figures sent as eps
files.

We would appreciate the publication of our paper.

Sincerely Yours,

R.J. Radwanski and Z. Ropka \\
Attachments: C:$\backslash$3d-03bis$\backslash$d1-V4plus$\backslash$Radwanski-Na2V3O7x.tex\\
C:$\backslash$3d-03bis$\backslash$d1-V4plus$\backslash$Fig-1.eps;
C:$\backslash$3d-03bis$\backslash$d1-V4plus$\backslash$Fig-2.eps;\\
C:$\backslash$3d-03bis$\backslash$d1-V4plus$\backslash$Fig-3.eps;
C:$\backslash$3d-03bis$\backslash$d1-V4plus$\backslash$Fig-4.eps.\\
\begin{center}
\textbf{APPENDIX B}\\
\textbf{\small{Rejection, due to unsuitability, the publication in Phys.Rev.Lett. 26-08-2003}}
\end{center}
\bigskip
Due to reasons independent on the authors it is only available on www.css-physics.edu.pl
\bigskip
\begin{center}
\textbf{APPENDIX B1}\\
\textbf{\small{Report of Referee A -- LG9202/Ropka 17-08-2003 sent 26-08-2003 attached to APPENDIX B}}
\end{center}
\bigskip
Due to reasons independent on the authors it is only available on www.css-physics.edu.pl
\newpage
\begin{center}
\textbf{APPENDIX C}\\
\textbf{\small{Answer to the referee report A and RESUBMISSION 27-08-2003}}
\end{center}
From: sfradwan@cyf-kr.edu.pl \\
To: Physical Review Letters $<$prl@ridge.aps.org$>$\\
Subject: Resub PRL LG9202 Ropka \\
$>$ From : R.J. Radwanski and Z. Ropka Krakow, 27-08-2003

Answer to the referee report A and RESUBMISSION

$>$ concerns LG9202: Origin of large reduction of the effective moment in Na$_{2} $V$_{3}$O$_{7}$

Dear Editor of Phys. Rev. Lett.

Thank you very much for your email of 26-08-03 despite of the negative decision with respect to our paper LG9202. We
appreciate your efforts for searching for referees as we could see inspecting the computer query. We see that you did
not get reports from 2 referees despite of 5 weeks waiting - please insist on them to get these reports. Your negative
decision is based on the only one referee report (A). We read this report carefully and have to say that in this report
we do not find any serious scientific objections.

We answer to the report and we resubmit our paper. The paper has been slightly improved mainly with respect to English.
The paper accounts 6 pages, 8 references, 4 figures (the same, so they are not sent again).

Thus we resubmit our paper. We would appreciate its publications in Phys. Rev. Lett. owing to the fact that it provides
novel alternative explanation for phenomena published in Phys.Rev. Lett. {\bf 90} (2003) 167202.Writing a new paper
seems to be better than writing Comments.

Sincerely Yours,

R.J. Radwanski

PS. We expect that you can get these two other reports. Please let me know them as soon as possible.

2. The answer to referee report A is attached below. (APPENDIX C1)
\newpage
\begin{center}
\textbf{APPENDIX C1}\\
\textbf{\small{Answer of R.J.Radwanski and Z. Ropka to the referee report A 27-08-2003}}
\end{center}
$>$concerns LG9202: Origin of large reduction of the effective moment in Na$_{2} $V$_{3}$O$_{7}$\\ $>$by Z. Ropka and
R.J. Radwanski

We have read this report carefully and have to say that in this report we do not find any serious scientific
objections. The refuse of the recommendation by the referee is hardly understood. noting that:

1. The referee admits, paragraph 1, that our paper gives novel explanation for temp. dependence of paramagnetic
susceptibility chi(T) of Na$_{2}$V$_{3}$O$_{7}$,

2. The referee admits, at the end of paragraph 1, that we can well explain the temperature dependence chi(T). Moreover,
he admits (beginning of paragraph 2) that our fit to the exp. data is quite good (in physics it is very high opinion).

3. the referee writes, the beginning of paragraph 2, that the good fit ''may indicate that the present interpretation
is valid'' and refuses his recommendation.

We hardly can understand the meaning of ''What is not clear is the significance and the magnitude of the resulting
crystal field parameters and the spin-orbit coupling constant.'' (paragraph 2).

Everybody knows that in solids, containing transition-metal atoms, there are crystal field interactions, discrete
states and that there is spin-orbit coupling. Should we take this sentence that the referee questions the existence of
these well-known in physics phenomena. If so, than we openly disagree with the referee. Not referring to our earlier
papers we can refer here to, e.g., just found Phys.Rev. B paper {\bf 65} (2002) 212405 about LaTiO$_{3}$. In Fig. 1 in
the inset there is a schematic electronic structure just as our presented in Fig. 1 (3). But we have to emphasize
that, in contrary to the schematic picture like that presented in PRB {\bf 65%
} (02) 212405, we have calculated our electronic structure and the resulting properties.

The only clearly-defined objections in paragraph 3, that

1. ''it is not clear whether assumptions of the calculations make sense'', we answer that our calculations are based on
well-known concepts in physics that have been successfully verified in, e.g., rare-earth compounds; secondly - the best
proof is the power in description of properties of real compounds,

2. the existence of local distortions should be demonstrated - we answer that in Na$_{2}$V$_{3}$O$_{7}$ the local
distortions from the octahedral symmetry is well known - the local symmetry of the V$^{4+}$ ion is much lower than
octahedral (Ref. 1).

Concluding, our approach to Na$_{2}$V$_{3}$O$_{7}$ is in the general line of the present understandings of magnetic and
electronic properties of transition-metal compounds. We have calculated the chi(T) dependence using well-known physical
concepts. We hope that the referee is satisfied with our answer and we ask him for recommendation of our paper.\\

\begin{center}
\textbf{APPENDIX D}\\
\textbf{\small{Rejection, due to unsuitability, the publication in Phys. Rev. Lett. 04-09-2003}}
\end{center}
\bigskip
Due to reasons independent on the authors it is only available on www.css-physics.edu.pl
\bigskip
\begin{center}
\textbf{APPENDIX D1}\\
\textbf{\small{Report of Referee B -- LG9202/Ropka - sent by email 4-09-2003 attached to APPENDIX\ D}}
\end{center}
\bigskip
Due to reasons independent on the authors it is only available on www.css-physics.edu.pl
\bigskip
\begin{center}
\textbf{APPENDIX E}\\
\textbf{\small{Answer to the referee report B and RESUBMISSION 11-09-2003}}
\end{center}
To: \quad prltex@aps.org \\
Subject: \quad resub PRL LG9202 Ropka\\
Date sent: \quad Thu, 11 Sep 2003 18:11:10 +0200\\
From:\quad R.J. Radwanski and Z. Ropka Krakow, 11 Sept 2003\\

Answer to the referee report B and RESUBMISSION

concerns LG9202: Spin-orbit origin of large reduction of the effective moment in Na$_{2}$V$_{3}$O$_{7}$

Dear Editor of \ Phys.Rev. Lett.

Thank you very much for your email of Sept 4, 2003 despite of the negative decision with respect to our paper LG9202.

We read two referee reports A (26 August) and B (4 Sept) carefully and have to say that in these reports we do not find
any serious scientific objections.27-08 we answer to the report A - here we answer to Report B and we resubmit our
paper asking The Editor of Phys. Rev. Lett. for reconsideration of his negative decision of 4-9-2003. Our paper offers
a
novel original solution to the phenomena published in Phys. Rev. Lett. {\bf %
90} (2003) 167202. We present an alternative explanation, that is more physical. It indicates that our explanation is
much, much superior.

We are convinced that it is better to present this novel explanation in a new paper than to write Comment. Thinking in
this way we RESUBMIT our paper. The paper has been slightly improved mainly with respect to English. Title has been
slightly changed. The paper accounts 5 pages, 11 references, 4 figures (the same as originally submitted 10 July 2003).
3 own references have been added that could help referees to understand our paper though we think that referee
understand our paper but there is simply conflict of interest. They cannot accept that physics can be so simple and
that the role of spin-orbit coupling can be so important in 3d-atom compounds.

Being convinced that for Science the open discussion and free exchange of information is fundamentally important we
resubmit our paper. We would appreciate its publications in Phys. Rev. Lett..

Sincerely Yours,

R.J. Radwanski

PS. We attach answer to referee report B. (APPENDIX\ E1). We attach also our answer to referee report A sent
27-08-2003. (APPENDIX C1).
\bigskip
\begin{center}
\textbf{APPENDIX E1}\\
\textbf{\small{Answer of R.J.Radwanski and Z. Ropka to the referee report B of 4-9-2003\\and RESUBMISSION 11-09-2003}}
\end{center}
$>$concerns LG9202: Origin of large reduction of the effective moment in Na$%
_{2} $V$_{3}$O$_{7}$

We have read your report carefully and have to say that in this report we do not find any serious scientific
objections. The refuse of the recommendation by the referee is hardly understood, not saying that it is scientifically
dishonest.

1. The referee B writes ''The deviation from the experimental results seems significant and must be explained.'' - it
is fully wrong as our calculated line 5 follows exactly the experimental data shown as x in Figs 2,3,4. Thus referee
must be wrong.

2. The referee B writes ''A consistency of a set of parameters B$_{4}$, B$%
_{2}$ .... must be checked to compare ...'' - we use only three parameters B$%
_{4}$, B$_{2}$ and lambda and all of them have clear physical meaning (this sentence has been now added to the paper).
Also we compare the used value of B$_{4}$=+200 K to that found recently experimentally in LaCoO$_{3}$ (also we put it
into paper: in Conclusion). Due to this we had to add 3 our references.

3. The referee B writes :''The paper leaves much room...'' we understand as plus for our paper. It proves the
prediction power of our approach. So, how you can in such situation refuse the recommendation. You could write that the
paper is scientifically important but should be improved.

Moreover, I suppose that you agree that our paper should be published in Phys. Rev. Lett. owing to the fact that it
offers novel explanation for phenomena published in Phys. Rev. Lett. {\bf 90} (2003) 167202.

We hope that the referee is satisfied with our answer and we ask you to recommend our paper for publication.

Sincerely Yours,

Z. Ropka and R.J. Radwanski

Attachments:

C:$\backslash$3d-03bis$\backslash$d1-V4plus$\backslash$Radwanski-Na2V3O7x91.tex
C:$\backslash$3d-03bis$\backslash$d1-V4plus$\backslash$Fig-1.eps;
C:$\backslash$3d-03bis$\backslash$d1-V4plus$\backslash$Fig-2.eps;
C:$\backslash$3d-03bis$\backslash$d1-V4plus$\backslash$Fig-3.eps;
C:$\backslash$3d-03bis$\backslash$d1-V4plus$\backslash$Fig-4.eps;
\bigskip
\begin{center}
\textbf{APPENDIX F}\\
\textbf{\small{Appeal to DAE and the challenge of Phys. Rev. Lett. referees 11-09-2003}}
\end{center}

To: \qquad prl@aps.org

Subject: \qquad Editor PRL - status LG9202 Ropka

Copies to: \qquad franz@aps.org, blume@aps.org

Date sent: \qquad Mon, 15 Sep 2003 18:11:21 +0200

From : R.J. Radwanski and Z. Ropka Krakow, 15 Sept 2003

concerns LG9202: Spin-orbit origin of large reduction of the effective moment in Na$_{2}$V$_{3}$O$_{7}$:

Dear Editor of Phys. Rev. Lett.

11 Sept. 2003 we have resubmitted our paper with the kind request to reconsider the negative decision from 9-04-2003.
Otherwise, please consider this letter as APPEAL to DAE.

We have answered to two referee reports A and B.

We consider these reports as not scientifically honest.

We know that the Phys. Rev. Lett. has the best-experienced referees, but we challenge PRL referees. We are ready for
the scientific open discussion.The referee reports prove that our explanation and approach is novel at present days,
though it makes use of the crystal-field interactions. In contrary to most, if not all, present papers we take into
account the relativistic spin-orbit interactions. In such situation the rejection decision of the Editor violates the
scientific rules, the more that our paper offers novel description to phenomena described in Phys. Rev.Lett paper {\bf
90} (2003) 167202).

We chose the writing of paper instead of Comments$^\ast$.

Being convinced that for Science the open discussion and free exchange of information is fundamentally important we ask
to publish our paper in Phys. Rev. Lett. If it could remove obstacles we agree to publish it together with the negative
referee reports.

Sincerely Yours,

Z. Ropka and R.J. Radwanski

Information to:

1. President of Amer. Phys. Society APS (office)

2. Editor-in-Chief dr M. Blume \\
\line(10,0){470} \\
$^\ast$ - Note added 25-Nov-2003. The Comment, due to the rejection of 16 October 2003, has been submitted 31 October
2003 as LKK921 to Phys. Rev.Lett.. See also cond-mat/0311033. See also our new paper on this topic cond-mat/0311575.

\end{document}